\def\etal{{\it et al.\/}}

\def\ie{{\it i.e.\/}}

\def\Mesz{{M\'esz\'aros}}

\documentstyle[aaspp4]{article}
\tighten

\begin{document}

\title{The soft X--ray afterglow of gamma ray bursts, \\
a stringent test for the fireball model}
\author{Mario Vietri}
\affil{ Osservatorio Astronomico di Roma \\ 00040 Monte Porzio Catone (Roma),
Italy \\ E--mail: vietri@coma.mporzio.astro.it \\
}

\begin{abstract}

I consider the recent discovery of a soft X--ray source inside the error box
of the gamma ray burst GB 960720 by the SAX, ASCA and ROSAT satellites, in terms
of the fireball model. I show that the ejecta shell, which, after 
causing the burst is cold and dense, but still relativistic, keeps plowing
through the interstellar medium, heating up the just--shocked matter
which then emits X--rays. I compute the radiation emitted by this 
matter. I show that, up to about two months after the burst, in the
cosmological scenario a soft X--ray ($0.1-10 \; keV$)
flux of at least $\approx 10^{-13} \;
erg \; s^{-1} \; cm^{-2}$, well within current observational capabilities,
is generated, explaining the observations of the three satellites.
Instead, in the Galactic Halo scenario a flux $3$ orders
of magnitude lower is expected. Detection of this non--thermal, declining flux 
in a statistically significant number of objects would simultaneously establish 
the fireball model and the cosmological nature of gamma ray bursts. 

\end{abstract}

\section{Introduction}

Our current theoretical understanding of gamma ray bursts (GRBs) is 
mostly based upon the fireball model (\Mesz, Laguna, Rees 1993). 
However, though this model has won great critical acclaim because 
of its ability to explain two otherwise mysterious features (the
bursts' time duration and the nonthermal spectra), it still has
made no testable predictions that would allow gauging it against
observations. This is mostly due to our ignorance of the physics
of electron acceleration at relativistic shocks. To get a feeling
of how serious this problem is, notice that the very same, simplified
analysis of the radiation emitted behind the shock
was carried out in two papers treating exactly the same physical
problem, but separated by twenty years, Blandford and McKee (1977,
dealing with AGNs) and Sari, Narayan, Piran (1996, dealing with
GRBs), and that furthermore the first one was published before
particle acceleration at shocks was even discovered (Bell 1978). 

It is the aim of this paper to derive a prediction from the fireball
model, by (nearly completely) circumventing the problem of electron acceleration at 
relativistic shocks. Surprisingly, the prediction is different for 
cosmological and Galactic Halo scenarios. The opportunity to do this
is offered by the detection, by
the X--ray satellite SAX (Piro \etal, 1995), of the 
gamma ray burst GB960720, both in the hard X--ray/soft $\gamma$--ray
band, where its results are confirmed by simultaneous observations by
BATSE in a nearly identical band, and in the soft X--ray band (Piro
\etal, 1996a). Subsequent observations of GB 960720 (Piro \etal, 
1996b, Murakami \etal, 1996, Greiner \etal, 1996) made $\approx 45$ 
days after the burst, have shown that a weak source, which is not an AGN, 
is present inside the WFCs' error box, $\approx 5\; arcmin$, with a flux 
$\approx 2\times 10^{-13}\; erg\; s^{-1}\; cm^{-2}$ in the band $0.1-10 \; keV$.

It is thus interesting to speculate about what a systematic 
search for GBRs' afterglow ought to yield, in the fireball model. 
This afterglow has a simple interpretation: it is the
cooling of matter swept up by the (still relativistic) shell of ejecta,
plowing through the interstellar medium. 
In the next Section, I compute the expected soft X--ray fluxes from
GRBs some time after the burst, for the fireball model, both in the 
cosmological and in the Galactic Halo scenarios. In the last Section, 
I discuss how the theoretical computations relate to the SAX, ASCA and 
ROSAT observations, and why the soft X--ray band is ideal for carrying out 
a statistical search for a soft X--ray afterglow.

\section{Predictions for the cosmological and Galactic Halo scenarios}

In the fireball model (\Mesz, Laguna, Rees, 1993), an initial energy
$E$ is released, together with a contaminating mass of baryons
$M_{ej} = E/\eta c^2$, with $\eta \ga 10^2$, as required
by observations. In the cosmological scenario, $E = 10^{51} E_{51} \; erg$,
and $M_{ej} \ga 10^{28}\; g$. For reasons to be explained later, I shall
concentrate on long ($\ga 1\; s$) bursts, for which $\gamma \approx 100$
(Sari and Piran 1995). In this case, after
a phase of free expansion, when the expanding ejecta have swept
up an ISM mass $\approx M_{ej}/\eta \ll M_{ej}$, a shock with the
ISM forms, at a distance from the site of energy injection
\begin{equation}
\label{rshock}
R_{sh} = 
5\times 10^{16} n_1^{-1/3} E_{51}^{1/3} \left(\frac{100}{\eta}
\right)^{2/3} \; cm\;,
\end{equation}
where $n = 1\; n_1 \; cm^{-3}$ is the ISM density. At this point,
a reverse shock propagates backward toward the still freely expanding ejecta, 
converting their directed kinetic energy into internal energy, whose
prompt release (on a timescale of seconds) leads to the GRB. 

Conventional analysis halts here, but the shell still has 
a Lorenz factor $\eta/2$ (Sari and Piran 1995), so that it will 
continue to expand relativistically into the interstellar medium.
Post--shock material has cooling times of order $\approx 10^3\; s$ in
the shock frame (\Mesz, Rees, Laguna 1993, Sari, Narayan, Piran 1996), 
depending on various details, but always negligible with respect to
hydrodynamical evolution times of the post--burst shell which, as will
be shown later, are of order $\approx 1$ month. Thus, the shell can be regarded 
as cold and dense; its evolution is given by Blandford and McKee (1976) as
\begin{equation}
\label{gamma}
\gamma -1 = \frac{2}{ \frac{M^2}{M^2_{ej}} \frac{\eta +2}
{\eta-2} - 1} 
\end{equation}
where I used the fact that the shell has initial Lorenz factor $\eta/2$.
This equation is exact for any shock (and shell) speed, sub or super 
relativistic.
Here $M$ is the present mass in the shell, $M= M_{ej} + n m_p V$ where
$V$ is the volume swept up. From the above equation, we see that the 
shell is relativistic ($\gamma \ga 2$) until $M \approx \sqrt{3} M_{ej}$,
\ie, till it reaches a radius
\begin{equation}
\label{rrel}
R_{rel} \approx \eta^{1/3} R_{sh} = 
2.4\times 10^{17} n_1^{-1/3} E_{51}^{1/3} \left(\frac{100}{\eta}
\right)^{1/3} \; cm\;.
\end{equation}

As long as the expansion is relativistic, eq. \ref{gamma} can be rewritten, 
defining an adimensional time since the burst $x\equiv c t/R_{sh}$, as
\begin{equation}
\label{gammagood}
\gamma = 1 +\frac{2}{(1+\frac{(1+x)^3-1}{\eta})\frac{\eta+2}
{\eta-2}-1}\;.
\end{equation}
Since the cooling time is much shorter than the shock evolutionary time, 
the total energy radiated per unit time, a relativistic invariant,
also follows from hydrodynamical arguments (Blandford and McKee 1976):
\begin{equation}
\label{edot}
\dot{E} = 4\pi R^2 v \gamma (\gamma-1) n m_p c^2
\end{equation}
where $R, \gamma$ and $v$ are the shock position, Lorenz factor and speed,
respectively. It is convenient to rewrite Eq. \ref{edot} as
\begin{equation}
\label{edotgood}
\dot{E} = 4\pi R_{sh}^2 n m_p c^3 f(x) = 
10^{42} n_1^{1/3} E_{51}^{2/3} \left(\frac{100}{\eta}\right)^{4/3}
f(x) \; erg \; s^{-1} \;,
\end{equation}
where 
\begin{equation}
\label{fofx}
f(x) \equiv (1+x)^2 (\gamma-1)\sqrt{\gamma^2-1} \;.
\end{equation}
Because of the superluminal expansion effect, the time as measured by an
observer on Earth, $t_\oplus$, is given by $dt_\oplus = (1-v/c) dt$; I 
define an adimensional Earth time as 
\begin{equation}
\label{y}
d y \equiv \frac{c\;d t_\oplus}{R_{sh}} = (1 - 
\frac{\sqrt{\gamma^2-1}}{\gamma}) d x \;;
\end{equation}
Eqs. \ref{y} and \ref{fofx} together define parametrically the dependence
of the luminosity in terms of Earth time; this is plotted in Fig. 1, for
the range of time for which the shell is relativistic. From this (and
Eq. \ref{rrel}) we see that the shell is relativistic for $\approx 2 \; month$ 
after the burst. From Eq. \ref{rrel}, it can be seen that the total
distance (and thus the time) before slowdown to sub--relativistic speed
is only mildly dependent upon $\eta$
(like $\eta^{1/3}$), so that the total lapse of time before afterglow turnoff 
is reasonably well determined  and, as pointed out to me by the referee, 
roughly the same for both short and long bursts, which are thought to differ
mostly because of their different $\eta$--values. 
Also, from Fig. 1 it can be seen that $f(y)$ is a curve with 
some curvature, so that no power--law dependence of $\dot{E}$ on time $t$ is 
meaningful. Lastly, notice that the exact shape of Fig. 1 is somewhat
sensitive to $\eta$, though not its normalization after
$\approx 1\; month$. 

In the cosmological scenario, for a typical source distance $D = 1\; Gpc$,
I obtain for the total flux radiated by the shell in the relativistic 
snowplow phase, from Eq. \ref{edotgood}
\begin{equation}
\label{flux}
F = 10^{-14} \; erg \; s^{-1} \; cm^{-2} 
n_1^{1/3} E_{51}^{2/3} \left(\frac{100}{\eta}\right)^{4/3} 
\left(\frac{1\; Gpc}{D}\right)^2 f(y)
\end{equation}
from which we see that, in the whole first month after the burst, the total
flux level always exceeds $\approx 3\times 10^{-13} \; erg \; s^{-1} \; cm^{2}$.

Before estimating which fraction of this flux ends up in the soft X--ray
regime, I derive the total flux in the Galactic Halo scenario. Since
in this case sources are $\approx 10^4$ times nearer, the total energy 
released is $10^8$ times lower, $E= 10^{43} \; erg$. Since approximately
the same value of $\eta = 10^2$ is required both by burst duration and 
by the spectra, the contaminating baryon mass is now $M_{ej} = E/\eta
c^2 = 10^{20} \; g$  (Begelman, \Mesz, and Rees, 1993). In this scenario, a
reasonable ISM density is $n_1 \approx 10^{-3}$.

To estimate the total luminosity after about one month,
I find that the total distance covered by the shell, if it were still
relativistic, would be $R = 10^{17} \; cm$, containing a mass $M_{sw} = 
2n_1\times10^{27} \; g$. From Eq. Eq. \ref{gamma} we see that the shell
cannot be relativistic, because $M_{sw}^2/M_{ej}^2 \approx n_1^2 10^{14} \gg 1$.

Rather than integrating Eq. \ref{gamma} exactly, the following simple 
argument will be used. The shell ceases to be relativistic for 
$R_{rel} \approx \eta^{1/3} R_{sh}$ as before,
and, from Eq. \ref{rshock}, I find $R_{rel} \approx 1.1\times 10^{15} \; cm$.
This occurs $3.5\times10^4\; s$ after the burst. From that moment on,
the expansion is subrelativistic and the distinction between $t$ and 
$t_\oplus$ ceases to be important. In this limit, Eq. \ref{gamma}
becomes (Blandford and McKee 1976)
\begin{equation}
M v = 2 M_{ej} c \left(\frac{\eta-2}{\eta+2}\right)^{1/2}
\approx 2 M_{ej} c \;,
\end{equation}
which is the well--known nonrelativistic snowplow model, for which 
the shock position and speed scale as $R\propto t^{1/4}$, $v \propto
t^{-3/4}$. Using as initial values the previously determined position and 
time at which the shell becomes subrelativistic, I find that, a month
after the burst, the shock speed is $v \approx 10^9 \; cm 
\; s^{-1}$, and $R=3.5\times 10^{15} \; cm$. With these values the total flux 
emitted can be computed from the subrelativistic limit of Eq. \ref{edot},
$\dot{E} = 4\pi R^2 v^3 n m_p$. Using a typical source distance of $100
\; kpc$, I find
\begin{equation}
F = 1.\times 10^{-16} \; erg\; s^{-1}\; cm^{-2} \left(\frac{3\times10^6\;s}{t}
\right)^{7/4}\left(\frac{100\; kpc}{D}\right)^2 \left(\frac{10^{-3} \;
cm^{-3}}{n}\right)^{1/4}
\end{equation}
for the Galactic Halo scenario. A month after the burst, the flux is three
orders of magnitude below that of the cosmological scenario, and, even if it
were totally emitted in the soft X--ray band, totally
unaccessible to current observational apparata. 
Thus an easy discrimination between the two models is possible.

I now estimate the fraction of all flux radiated in the soft X--ray band,
in the cosmological scenario. To do so, I have to resort to the
customary treatment (Blandford and McKee 1977, \Mesz, Laguna and Rees
1993, Sari, Narayan and Piran 1996) of electrons behind relativistic
shocks. Thus I shall suppose that about half of the total internal
energy is in a power--law distribution of electrons with index
$p=2.5$, giving rise to a synchrotron spectrum of index $q=(p-1)/2=
0.75$  (Band \etal, 1993, Sari, Narayan and Piran 1996). Inverse
Compton cooling does not contribute to emission in the relatively
low energy bands I am interested in, so that the fraction $f_X$
of the total energy emitted in a band with photon energy
$\epsilon_l < h\nu < \epsilon_u$ is
\begin{equation}
\label{fx}
f_X = \frac{\epsilon_u^{(3-p)/2} - \epsilon_l^{(3-p)/2}}{\epsilon_m^{(3-p)/2}}
\end{equation}
where $\epsilon_m$ is the energy, in the observer's frame,
of the typical bremsstrahlung photons emitted by the highest energy 
electrons (\ie, those having Lorenz factor $\gamma_m$ in the shell's frame):
\begin{equation}
\label{epsilonm}
\epsilon_m = \frac{\hbar e B}{m_e c} \gamma_m^2 \gamma\;;
\end{equation}
the Lorenz factor of the shell, $\gamma$, appears here to convert the
photons' energy from the shell's frame to the observer's. I estimate
$\gamma_m$ by equating, as usual, the electron acceleration timescale
with the synchrotron slow--down timescale, finding $\gamma_m \approx
10^7 B^{-1/2}$ (\Mesz, Laguna and Rees 1993). Inserting this
into Eq. \ref{epsilonm} shows that $\epsilon_m$ does not depend
upon the magnetic field, and so it is independent of another major
uncertainty of the problem, the efficiency with which equipartition
magnetic fields are built up behind a relativistic shock. From
Eq. \ref{fx} and Eq. \ref{epsilonm}, in the band $0.1-10\; keV$, for
$p=2.5$, I find
\begin{equation}
\label{fx2}
f_X \approx \frac{0.2}{\gamma^{1/4}}\;.
\end{equation}
Thus, the expected soft X--ray flux is
\begin{equation}
\label{flux2}
F_X = 10^{-14} \; erg \; s^{-1} \; cm^{-2} 
n_1^{1/3} E_{51}^{2/3} \left(\frac{100}{\eta}\right)^{4/3} 
\left(\frac{1\; Gpc}{D}\right)^2 f(y) f_X
\end{equation}
where the function $f(y) f_X$ is also plotted in Fig. 1. The soft X--ray flux 
is of order $\approx 10^{-13} \; erg \; s^{-1} \; cm^{-2}$. The above equation 
shows why I concentrated on long bursts: short bursts having $\eta \approx 
10^3$ (Sari and Piran 1995) have soft X--ray fluxes which are lower
than long bursts by about a factor of $20$. This makes them unobservable
with current satellites, even though still brighter than in the 
Galactic Halo model. 

A major uncertainty in estimating $f_X$ lies in postulating that the 
nonthermal electron population absorbs about half of all internal energy 
generated at the shock, but uncertainty in the slope of the spectrum 
also contributes. For $p=2$, in fact, I would have found $f_X \approx 0.1
/\gamma^{1/2}$, while, using $\gamma_m = 10^8$, would have resulted in
$f_X \approx 0.01 /\gamma^{1/2}$. 

By analogy with supernova remnants, as the shock slows down, the fraction of 
total internal energy absorbed by the nonthermal electron distribution
probably decreases, and becomes even harder to estimate. For this reason the
theoretical prediction has not been extended to subrelativistic speeds of
the shock.

\section{Discussion}

The most promising way to attack this problem is to 
try to identify afterglow emission from GRBs' error box
in the soft X--ray on a statistical basis.
The advantage of doing this in the soft X--ray rather than
in lower energy bands, where source contamination is also low, is that the
expected fluxes from Eq. \ref{fx} can be seen to be at least an order of
magnitude higher. Higher--energy bands, instead, do not have the
angular resolution necessary to avoid source confusion.
In the radio band, where lower fluxes are offset by much larger collecting
areas, a behaviour similar to that of the soft X--ray band is of course
expected, but theoretical computation is made difficult by several
subtleties. Consideration of this effect is thus postponed to a
forthcoming paper.

The small angular resolution of the WFCs onboard SAX ($5\; arcmin$)
allows follow up 
observations by narrow field instruments with the hope of little source 
confusion. This is exactly what has been done by Greiner \etal (1996),
who identified three sources with the ROSAT HR Imager inside SAX error
box. Of these, two sources are AGNs, while the third one, accounting for
about half of the total flux (\ie, $\approx 2\times 10^{-13} \; erg\; s^{-1}
\; cm^{-2}$) detected by SAX and ASCA, has no optical 
counterpart, indicating a very unusual object.

From Fig. 1, where the tickmark indicates the position of $43$ post--burst
days, and Eq. \ref{flux2}, we see that the expected flux, $5\times10^{-14}
\; erg\; s^{-1}\; cm^{-2}$, compares remarkably well (and perhaps fortuitously,
since we do not know the source distance) with the observation of 
$10^{-13} \;erg\; s^{-1}\; cm^{-2}$ (Piro \etal, 1996b, 
Murakami \etal, 1996, Greiner \etal, 1996). Also,
it should be noticed that Piro \etal\/(1996a) have set an upper limit
to the soft X--ray flux from the burst region, immediately after the 
burst, of $10^{-10} \; erg \; s^{-1} \; cm^{-2}$. Using Eq. \ref{flux2}
and Fig. 1 it can be seen that the highest flux expected in this model
is $\la 10^{-11} \; erg \; s^{-1} \; cm^{-2}$. Thus, the current model
reproduces correctly, for the most trivial choice of parameters, 
the observed features of the afterglow of GB960720. Hopefully, 
future observations ought to show that the afterglow has disappeared
on a timescale of a few months, even though for this source it will be 
impossible to determine whether the light curve follows Fig. 1.

Also, the apparent lack of interstellar absorption is consistent with the
fireball model. The total baryon contamination $\approx 10^{27} g$ (times
at most a factor of $2$ to include the mass swept up when the shell is still 
relativistic) when spread out over a spherical surface of radius 
$R_{sh} < R < R_{rel}$ (Eqs. \ref{rshock} and \ref{rrel}) provides a
total column depth $N_H \approx 10^{16} - 10^{17} \; cm^{-2}$, well within
observational constraints. If furthermore, as has been suggested, GRBs
are related to mergers of neutron star binaries, they are expected to be
distributed somewhat like pulsars, \ie\/ outside the galactic disk, where 
column depths do not approach the observational bounds.

Lastly, I would like to point out the reason why detection of soft 
X--rays about a month after the burst can discriminate between 
cosmological and Galactic Halo models, since it is rather amusing.
In the fireball model, there are $5$ dimensional 
parameters: $E, M_{ej}, \rho_{ISM}, c, D$, where $\rho_{ISM}$ is the density 
of the circumstellar matter, and $D$ is source distance. Of these, $c$ is
a universal constant, and $\rho_{ISM}$ is an external parameter, which ought
to be considered as given, and which is, furthermore, relatively well--known,
compared with the uncertainty in parameters such as $E, M_{ej}, D$, each
spanning several orders of magnitude. We can thus regard it as
fixed. Thus, specifying that the two observational constraints,
flux and time duration at Earth, be reproduced means fixing $2$ of the
$3$ free parameters, leaving only one (say, $D$) undetermined. This
corresponds to having a one--parameter ($D$) family of homologous
solutions, each fitting observational data, each located at different 
distances from the observer. By making observations
at a fixed time after the burst, about a month, we are observing
cosmological and Galactic Halo models at non--homologous moments,
thus breaking the similarity law that links them: in fact, the shock
is still relativistic in the cosmological scenario, and well subrelativistic in 
the Galactic Halo scenario.

In short, I have argued that, in the cosmological scenario of the
fireball model, detectable fluxes (Eq. \ref{flux2} and Fig. 1) of soft 
X--rays should be emitted in the two months following a gamma ray burst,
with a non--thermal spectrum and a characteristic decrease (Fig. 1),
while no detectable flux can arise in the Galactic Halo scenario. 
In particular, the expected flux of $10^{-13} \; erg\; s^{-1} \; cm^{-2}$
compares remarkably well with the observations of GB960720 made
about $40$ days after the burst, by Piro \etal (1996b), Murakami \etal, (1996)
and Greiner \etal (1996). But the model also predicts the time--dependence of 
the afterglow (Fig. 1) and its disappearance after a few months, and is thus 
subject to more elaborate testing. Detection of the afterglow in a statistically
meaningful sample of gamma ray bursts would simultaneously establish both the
fireball model {\bf and} the cosmological nature of GRBs. 

\vskip 1truecm
I am indebted to L. Stella and especially to Luigi Piro, for fruitful 
scientific conversations, and to an anonymous referee for constructive 
criticisms.

\vskip 2truecm
\begin{figure}
\caption{
Solid line: plot of $f(y)$ (from Eq. \ref{fofx}), against adimensional time $c 
t_\oplus
/R_{sh}$ where $R_{sh}$ is defined in Eq. \ref{rshock}, with
$\eta = 100$. Dashed line: plot of the product
$f(y) f_X$, where $f_X$ comes from Eq. \ref{fx2}. The tickmark represents,
in scaled units, $43$ days after the burst, the moment at which the follow
up observations of GB 960720 were carried out by Piro \etal\/ (1996b). 
}
\end{figure}


\begin{references}
\reference{} Band, D., \etal, 1993, \apj, 413, 281.
\reference{} Begelman, M.C., M\'esz\'aros, P., Rees, M.J., 1993,
\mnras, 265, L13.
\reference{} Bell, A.R., 1978, \mnras, 182, 145.
\reference{} Blandford, R.D., McKee, C.F., 1976, Phys. Fluids, 19,
1130.
\reference{} Blandford, R.D., McKee, C.F., 1977, \mnras, 180, 343.
\reference{} Greiner, J., Hagen, H.J., Heines, A., 1996, IAU Circ. no. 6487.
\reference{} M\'esz\'aros, P., Laguna, P., Rees, M.J., 1993, \apj, 415, 181.
\reference{} Murakami, T., \etal, 1996, IAU Circ. no. 6481. 
\reference{} Piro, L., Scarsi, L., Butler, R.C., 1995, in {\it X--ray and
EUV/FUV Spectroscopy and Polarimetry}, S. Fineschi ed., SPIE 2517, 169.
\reference{} Piro, L. {\it et al.}, 1996a, \nat, submitted.
\reference{} Piro, L. {\it et al.}, 1996b, IAU Circ. no. 6480.
\reference{} Sari, R., Narayan, R., Piran, T., 1996, \apj, submitted.
\reference{} Sari, R., Piran, T., 1995, \apjl, 455, L143.
\end{references}
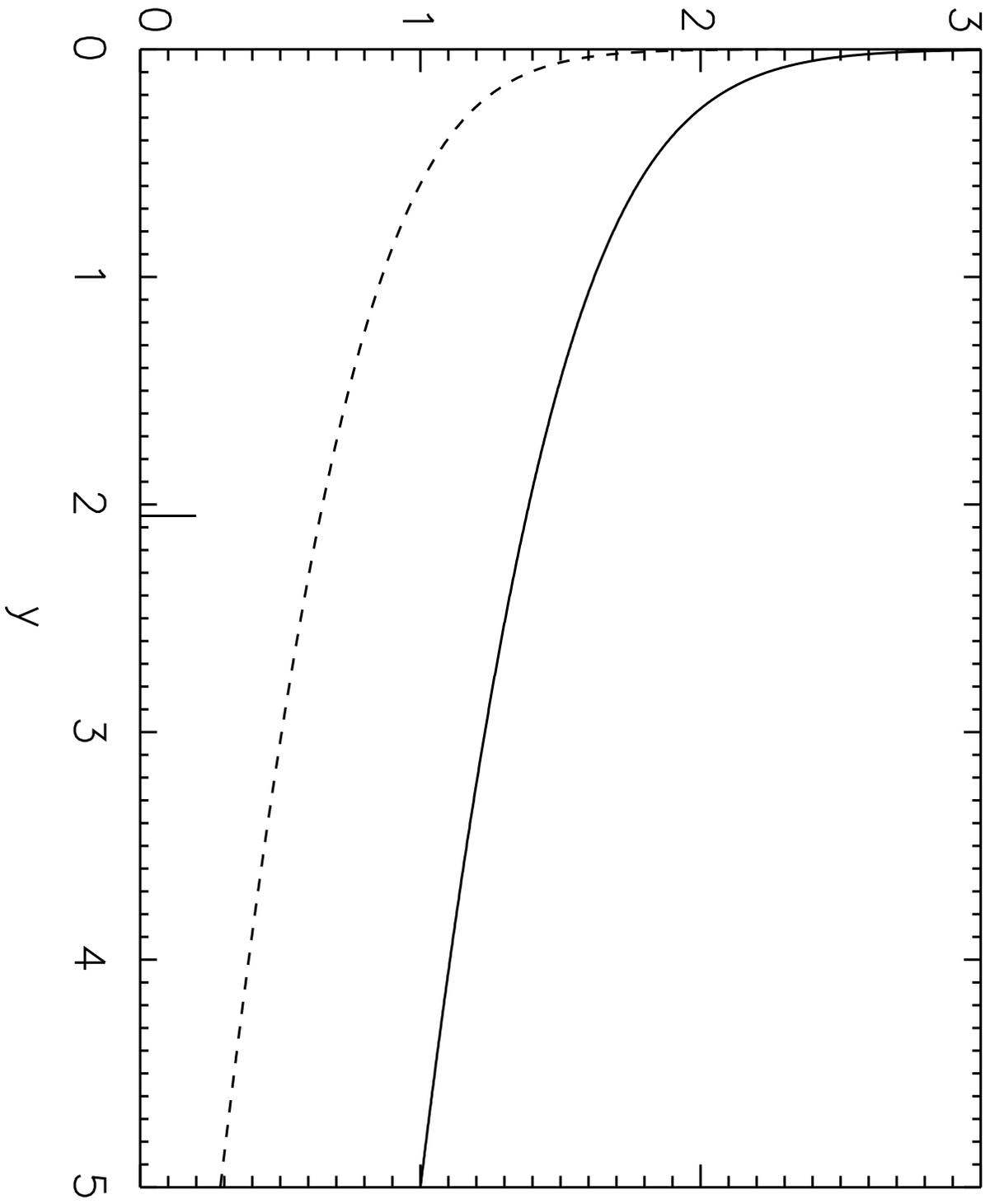
\end{document}